\documentclass[pra,twocolumn,superscriptaddress,showpacs]{revtex4}

\usepackage{graphicx,amssymb,amsfonts,latexsym,color,dcolumn,bm,epsfig}

\newcommand{\ket}[1]{\left| #1 \right\rangle}
\newcommand{\bra}[1]{\left\langle #1\right |}

\begin{document}

\title{Phase Dynamics of Entangled Qubits}
\author{P\'erola Milman}
\email{Perola.Milman@ppm.u-psud.fr}
\affiliation{Laboratoire de Photophysique Mol\'eculaire du CNRS, Universit\'e Paris-Sud, B\^atiment 210--Campus d'Orsay, \\
91405 Orsay Cedex, France}

\date{\today}


\begin{abstract}
We make a geometric study of the  phases acquired by a general pure bipartite two level system  after a cyclic unitary evolution. The geometric representation of the two particle Hilbert space makes use of Hopf fibrations. It allows for a simple description of the dynamics of the entangled state's phase during the whole evolution. The  global phase after a cyclic evolution is always an entire multiple of $\pi$ for  all bipartite states, a result that does not depend on the degree of entanglement.  There are three different types of phases combining themselves so as to result in the $n \pi$ global phase. They can be identified as dynamical, geometrical and topological. Each one of them can be easily identified using the presented  geometric description. The interplay between them depends on the initial state and on its trajectory and the results obtained are shown to be in connection to those on mixed states phases. 
   \end{abstract}

\maketitle
\section{Introduction}

Quantum phases are a peculiarity of quantum physics giving it some of its intriguing and unusual effects, such as interference. While relative phases are extremely important for any measurable property of a given quantum system, global phases are, in general, irrelevant and play no major part. Nevertheless, they can be measured using conditional dynamics. Supose one has an initial state described by  $1/\sqrt{2}\ket{\Psi}_1(\ket{0}_2+\ket{1}_2)$, where the subscripts refer to two different subspaces. Suppose now that, if particle $2$ is in state $\ket{1}_2$, $\ket{\Psi}_1$ gains a phase of $\phi$. By  rotating the second particle so as $\ket{1} \rightarrow \frac{1}{\sqrt{2}}(\ket{1}+\ket{0})$ and $\ket{0} \rightarrow \frac{1}{\sqrt{2}}(\ket{0}-\ket{1})$, we have that our state becomes $\frac{1}{2}((e^{i \phi}+1)(\ket{\Psi}_1\ket{0}_2+(e^{i \phi}-1)(\ket{\Psi}_1\ket{1}_2)$. Detecting the probability of finding the second particle in state $\ket{1}_2$ gives us $P_2 \propto 1- \cos{\phi}$ allowing for the determination of $\phi$.  

Based on this discussion, we can now define the total phase gained by a system after its evolution. The definition used here is the same as proposed by Pancharatnam \cite{PAN}. In his seminal work, he defined the phase of a quantum state $\ket{\Phi}$ relative to another state $\ket{\Psi}$ as
\begin{equation}\label{pancha}
\phi_t={\rm arg}\langle{\Psi}|{\Phi}\rangle.
\end{equation} 
This definition is called from now on the {\it total} (global) phase. The total global phase gain may  originate from a combination of phase effects of different types. One possible origin of the global phase gain is of dynamical nature. This phase is given by the eigenvalues of the Hamiltonian, determining the time evolution of its eigenstates. However, dynamical phases are not the only type of phase appearing during a state's evolution.  In order to be more precise, let's restrict ourselves to a  two level system, or a qubit. The Hilbert space of such a system has a certain geometry that can be described by the Bloch sphere. This geometry accounts for {\it geometric phase} gains. Such phases were first introduced in quantum mechanical systems by M. Berry  \cite{BERRY,BOOK} and can lead to some  amazing and counter intuitive  effects associated to particles and their trajectories. One of the  most famous of such effects is the Aharanov-Bohm effect, in which  a measurable meaning is given  to the vector potential. The geometric phase depends on the trajectory realized by the qubit under the action of an evolution operator. For cyclic evolutions, it is given by half the solid angle of the closed path traced in the Bloch sphere, i.e., the area of the surface enclosed by the trajectory.

The development of quantum mechanics motivated new questions and problems concerning geometric phases. These problems involve the study of geometric phases in many particle entangled systems \cite{SJOQVIST:PRA00}, such as  twin photons \cite{KWIAT:PRL91} and Bose-Einstein condensates \cite{COND}. Local and non local aspects of  geometric phase have also been studied \cite{AHARONOV:PRL00}, as well as its behavior in dissipative systems \cite{MARCELO}. More recently, it was found that geometric phases can also be useful in quantum information theory.  It was shown that conditional quantum logics, a fundamental part to realize universal quantum computation, can be done using geometric phases \cite{DUAN:SCIENCE01}, a result already demonstrated experimentally \cite{WINE:NATURE03}. The interest of these so called topological gates is that they are more robust and naturally fault tolerant \cite{KITAEV:ANNALS03}. 

These recent results together with the fundamental interest of the subject, are a motivation for the present work: we study here the geometric representation of the phase dynamics of pure two-level bipartite systems with an arbitrary degree of entanglement. For this kind of system, a third possible type of global phase gain can be identified: a {\it topological} phase, which is  a consequence of the geometry of the entangled two-level system. This phase has been studied in \cite{EU:PRL03} for maximally entangled states (MES) and it is at the origin of singularities  appearing in the phase of MES during a cyclic evolution \cite{SJOQVIST:PRA00}. The topological origin of this phase can be made clear by applying the geometric representation of pure entangled qubits presented in \cite{MOSSERI:JPA01}. As shown in the present paper, the use of the same representation allows for a clear geometric picture showing the origin and interplay between the three mentioned different types of phases during the state's evolution. It is shown that, after a cyclic evolution, the combination of the phases presented above always lead to a global phase gain of an entire multiple of $\pi$. This result, already known and verified experimentally \cite{NEUTRONS} for a single qubit is recovered here for entangled qubits with an arbitrary degree of entanglement.  

The paper is organized as follows: we start by studying the case of one qubit and analogous systems, investigating their  geometric representation and phases acquired during a cyclic evolution.  We do so to set notations and notions for  the geometric representation and phase dynamics of two qubit systems, presented in Section III. Finally, Section IV is devoted to the study of some examples of evolutions leading to different types of phase gains for entangled systems. We finish the paper with some concluding remarks and a brief discussion on how the presented ideas could be tested experimentally.

\section{One Qubit: Geometry and Phases}

\subsection{Pure states}
\begin{figure}[h]

\center

\includegraphics*[width=2 in, keepaspectratio]{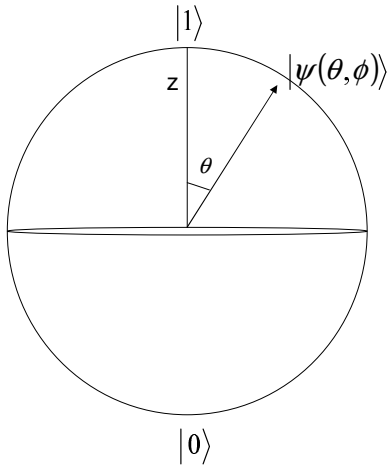}

\caption{Bloch sphere representation of an one qubit state.} \label{fig1}

\end{figure}

A general pure one qubit system can be represented by a $S^3$ sphere in $R^4$. $S^3$ Hopf fibrations define a map $S^3 \stackrel{S^1}{\rightarrow} S^2$, ``decomposing" the $S^3$ sphere into a $S^2$ base and a fiber. It has been shown \cite{URBANKE:AJP91,MOSSERI:JPA01} that this way of representing the space of one qubit is analogous to the Bloch sphere: the base can be identified to the Bloch sphere while the fiber represents the global phase indetermination of such a representation.   Taking the general pure state $\ket{\Psi(0)}=\alpha\ket{0}+\beta\ket{1}$, where $\alpha$ and $\beta$ are complex coefficients, its representation in  the Bloch sphere is shown in Fig. (\ref{fig1}). The state's coordinates can be obtained as follows:
\begin{eqnarray}
&&X=\langle \sigma_x \rangle_{\Psi}=2 \rm Re (\alpha^*\beta) \nonumber \\
&&Y=\langle \sigma_y \rangle_{\Psi}=2 \rm Im (\alpha^*\beta)  \\
&&Z=\langle \sigma_z \rangle_{\Psi}= |\alpha|^2-|\beta|^2 \nonumber. \\
\end{eqnarray}
As a consequence of the normalization condition, the radius of the Bloch sphere is equal to one. Rotations around axes in an arbitrary direction displace the state vector on the Bloch sphere. We can study, for example, the effect of a rotation around the $z$ axis, produced by the  Hamiltonian $\hat H=\hbar\omega \hat \sigma_z/2$. Physically, this may correspond to a magnetic field applied along the $z$ axis in the case where our qubits are spin $1/2$ particles. At time $t$, the state reads $\ket{\Psi(t)}=e^{\frac{-i\omega t}{2}}\alpha\ket{0}+e^{\frac{i\omega t}{2}}\beta\ket{1}$, where $\omega$ is the frequency of the applied field. In the Bloch sphere, this means that the state actually makes a precession around the $z$ axis, reaching again the initial state for $t=2\pi/\omega$.  This evolution leads to a phase gain. To compute the global phase gain of the state at any time $t$ and any point of the trajectory, one can use Pancharatnam's definition (\ref{pancha}). We see that after the $2 \pi$ rotation the global phase gain is equal to $\pi$, irrespectively of the choice of the initial state. Nevertheless, the {\it nature} of this phase depends on the initial state and on the Hamiltonian acting on it. Taking, for example, $|\alpha|=\cos{(\theta/2)}$ and $|\beta|=\sin{(\theta/2)}$, one can identify, in this simple one qubit case, a {\it dynamical} phase $\phi_d$, defined  as
\begin{equation}\label{dina}
\phi_d=\int_0^t \langle \hat U^{\dagger} \dot {\hat U}  \rangle dt,
\end{equation} 
where $\hat U$ is the evolution operator associated to the Hamiltonian $\hat H$. For the particular example studied, we have $\phi_d=-\pi \cos{\theta}$. As can be seen, there is a dependence on the angle $\theta$ that characterizes our initial state. We can also see that the dynamical phase does not account for the total phase gain. In order to get a complete description of the total phase gain, one should also consider the {\it geometric} phase $\phi_g$, which is a property of the space of states. In the case of qubits describing trajectories in  the Bloch sphere, the cyclic geometric phase corresponds to the area enclosed by the trajectory realized by the state vector on this same sphere. It assumes the value of $\phi_g=-\pi (1-\cos{\theta}) $ in our simple example. As can be easily checked, we have $\phi_t=\phi_g+\phi_d=-\pi$ after a cyclic evolution, and this result is state independent. The initial state determines only the proportion of the geometrical and dynamical phases.  

\subsection{Mixed states} 

\begin{figure}[h]

\center

\includegraphics*[width=1.8in, keepaspectratio]{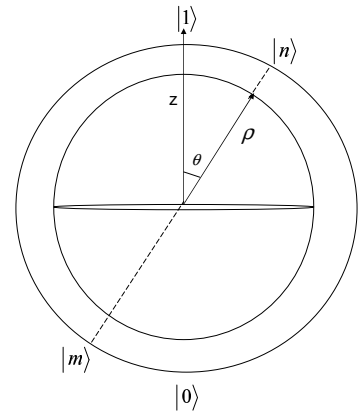}

\caption{Bloch ball representation of an one qubit density matrix and purification procedure.} \label{fig2}

\end{figure}

\begin{figure}[h]

\center

\includegraphics*[width=2.5in, keepaspectratio]{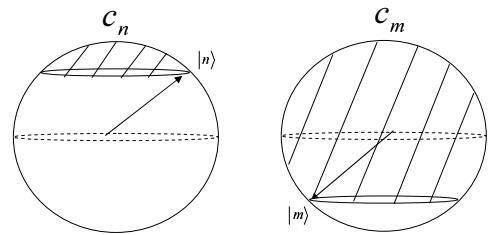}

\caption{geometric picture of the weighted sum of the trajectories corresponding to the extension of the density matrix vector to two pure states lying on the unitary Bloch ball. The areas considered are the marked ones. Coefficients $c_m$ and $c_n$ are the weights of each dashed area. } \label{fig3}

\end{figure}

One qubit mixed states are described by a density matrix of the form
\begin{equation}\label{dens1}
\hat \rho=a_{00}\ket{0}\bra{0}+a_{01}\ket{0}\bra{1}+a_{10}^*\ket{1}\bra{0}+a_{11}\ket{1}\bra{1}
\end{equation}
They can be geometrically represented in the Bloch ball, a generalization of the Bloch sphere. The Bloch ball has a radius smaller than one. It is, in fact, a function of the purity $P=1-\rm Tr \rho^2$ of the system, as follows: $r=|1-2P|^{1/2}$. $P$ assumes the value of $1/2$ for completely mixed states and $1$ for pure states. The coordinates of (\ref{dens1}) on the Bloch ball are given by
\begin{eqnarray}
&&X=\langle \sigma_x \rangle_{\Psi}=2 \rm Re (a_{01}) \nonumber \\
&&Y=\langle \sigma_y \rangle_{\Psi}=2 \rm Im (a_{01})  \\
&&Z=\langle \sigma_z \rangle_{\Psi}=a_{00}-a_{11} \nonumber. 
\end{eqnarray}
It is important to notice that the representation of a mixed state in the Bloch ball is different from the pure case. A point in the Bloch ball determines the density matrix with no global phase ambiguity. Also, the phases involved in a cyclic evolution of a mixed state deserves more attention. The global phase, analogous to the Pancharatnam's phase, is given by $\phi^M_t=\rm {arg Tr} [\hat U \rho(0)]$ \cite{OI:PRL00}. For the geometric phase, there are  different definitions in the literature \cite{OI:PRL00, TONG:PRA03, OI:PRL03}. The choice made here presents a clear advantage for our purposes, as will become clear in the following. In order to calculate the geometric phase, one should purify the mixed state appearing in the Bloch ball. This is done, geometrically, by prolongation of the density matrix vector until it reaches the unitary Bloch sphere. As a consequence, one gets two pure states, pointing in opposite directions (see Fig. (\ref{fig2})). The geometric phase is given by the weighted sum of each pure state's geometric phase. The weights appearing in this sum can be obtained from the density matrix as follows: one should diagonalize (\ref{dens1}) and express it in terms of the two weighted orthogonal components, i. e., $\hat \rho=c_m\ket{m}\bra{m}+c_n\ket{n}\bra{n}$, with $\langle m | n \rangle=0$. Coefficients $c_m$ and $c_m$ are the weights of the contribution of each state $\ket{m},\ket{n}$ to the geometric phase (see Fig. (\ref{fig3}). One possible interpretation of this writing of the density matrix is to suppose it comes from the trace of an entangled state written in the Schmidt's decomposition with respect to one of the qubits \cite{TONG:PRA03}.   

As a general rule, the total phase for mixed states is still a combination of a dynamical and a geometrical phase. As will be seen in more details in the following of this paper (Section IV), taking as an example a $2\pi$ cyclic evolution around a fixed axis, the sum of the geometrical and dynamical phases always leads to a total phase gain of $\pi$, as in the pure state case.


Rotations around one fixed axis studied above can lead to some ambiguous conclusions, as pointed out in \cite{EU:PRL03}: is the effect of $\pi$ dephasing a property of the two dimensional rotation group SU(2) or a property of general three dimensional rotations performed by the group SO(3)? Would the results of the experiment change if  the evolution operator acting on the qubits evolves in time? This corresponds to changing the magnetic field direction during the evolution in such a way that one  always have a cyclic evolution at the end. In order to answer these questions is to make use of MES. This was done in \cite{EU:PRL03} where  a particular geometric picture of the MES Hilbert space is used. More recently, some extensions to this study were made, always with the help of MES \cite{LIMING:PRA04}. In \cite{EU:PRL03} we made use of the geometric tools developed in \cite{MOSSERI:JPA01} to explicit the origin of the phase obtained by a MES in a cyclic evolution.  An interesting property of MES is that, as will become clear in the following, it does not gain, in a cyclic evolution, neither a geometric nor a dynamical phase. However, one can still observe a $\pi$ dephasing for some evolutions, identified as being from topological origin. In the following, a more complete geometric picture of the space of two entangled qubits is presented, and it is shown how different types of phases can appear for non MES.

\section{Geometric representation of two qubits}\label{remy}

In this section, we summarize the results presented in \cite{MOSSERI:JPA01} where a geometric representation of the space of pure two qubit states in presented. This is  done with the help of Hopf fibrations, as in the case of one qubit states. The main idea is to find a way which is analogous (or at least connected) to the Bloch sphere representation presented above. 

\subsection{Two Qubits Representation}
\begin{figure}[h]

\center

\includegraphics*[width=2.1in, keepaspectratio]{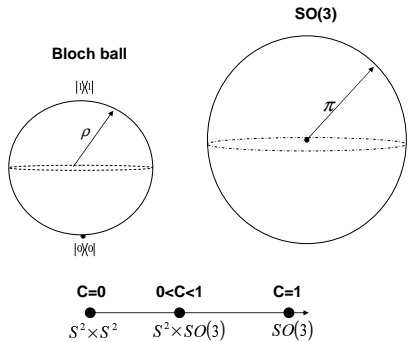}

\caption{Geometric representation of a pure two qubit system as a Bloch ball obtained from the trace with respect to one of the qubits and a sphere representing $SO(3)$. This representation is a function of the concurrence {$\cal C$} of the system. In the limit cases of MES (${\cal C}=1$) and product states (${\cal C}=0$), this representation reduces to $SO(3)$ or a product of two Bloch spheres, respectively. For a general non MES state ($0>{\cal C}>1$), the pure qubit state is represented by the Bloch ball and $SO(3)$ } \label{fig4}

\end{figure}

Consider a general pure entangled state with arbitrary coefficients $\alpha, \beta, \gamma$ and $\delta$:
\begin{equation}\label{initstate}
\ket{\psi_o}=\alpha\ket{00}+\beta\ket{01}+\gamma\ket{10}+\delta\ket{11}.
\end{equation}
State (\ref{initstate}) is normalized, and can be represented by a $S^7$ sphere in $R^8$. A tempting but insufficient way to represent (\ref{initstate}) geometrically is to make a partial trace of one of the qubits and study the properties of the resulting density matrix. This density matrix can be, as seen before, represented in the Bloch ball. This naive method, even if  dependent on the degree of entanglement of the initial state, lacks  information: the resulting mixed density matrix could  originate from an infinite number of entangled states with the same degree of entanglement. However, with the help of a map defining a Hopf fibration, one can combine to this reduced density matrix another space where the rest of the needed information can be found. This other space is the fiber space, defined as $S^3/Z_2=SO(3)$. The fiber space can  be represented as another $S^3$ sphere of radius $\pi$ with opposite points (points differing by a global phase) identified. The combination of the Bloch ball and the fiber completes the Hopf map, where $S^7$ is described by a basis $S^4$ and a fiber $S^3$ via the relation $S^7  \stackrel{S^3}{\rightarrow} S^4$ (see Fig (\ref{fig4})). It is the $S^4$ sphere's coordinates (base) which are related to the Bloch ball. They are given by
\begin{eqnarray}
&&X=\langle \sigma_x \rangle_{\Psi}=2 \rm Re (\alpha^*\gamma+\beta^*\delta) \nonumber \\
&&Y=\langle \sigma_y \rangle_{\Psi}=2 \rm Im (\alpha^*\gamma+\beta^*\delta)  \nonumber \\
&&Z=\langle \sigma_z \rangle_{\Psi}=|\alpha|^2+|\beta|^2-|\delta|^2-|\gamma|^2  \\
&&C_r= 2 \rm Re (\alpha\delta-\beta\gamma) \nonumber \\
&&C_i =2 \rm Im (\alpha\delta-\beta\gamma) \nonumber.
\end{eqnarray}  
The first three coordinates can be identified to the Bloch ball coordinates. The last two also appear in a way in the Bloch ball: they are connected to its radius as $r=\sqrt{1-C_r^2-C_i^2}$. Moreover, $C_r$ and $C_i$  are related to the concurrence $\cal C$ defined by Wooters \cite{WOOTERS:PRL98} by the relation ${\cal C}= |C_r + i C_i|=2|\alpha \delta - \beta \gamma|$. The concurrence is a measurement of the degree of entanglement for this system. It assumes the value of 1 for MES and 0 for separable states.

Some particularities of this representation that can be directly verified are worth being mentioned. For MES $({\cal C}=1)$, the Bloch ball reduces to a point.  MES are thus completely characterized by $SO(3)$. This  result is related to the fact that all maximally entangled states are connected by a local rotation. The reduction of MES to $SO(3)$ is also the reason why one cannot have geometric phases for MES. The global phase in this case should come only from the properties of $SO(3)$. For separable states $({\cal C}=0)$, the Bloch ball is the Bloch sphere, and the same happens to the fiber, so that one ends up with two independent Bloch spheres, one for each qubit, as expected. In the intermediate case, i.e., for states with $0<{\cal C}<1$, as mentioned before, one needs $SO(3)$ and the Bloch ball to completely represent the entangled state. In this case, {\it to each point in the Bloch ball is associated a fiber $SO(3)$}.

\begin{figure}[h]

\center

\includegraphics*[width=1.5in, keepaspectratio]{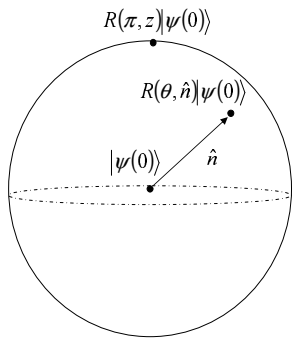}

\caption{A possible representation of $SO(3)$: a ball of radius $\pi$ with opposite points identified. Entangled states are points in this ball and displacements on it correspond to rotations applied to the entangled states. States in the border of the sphere are obtained by rotation of one of the qubits of the initial state (placed in the center of the sphere) of $\pi$.  } \label{fig5}

\end{figure}

Now we should gain some insight on how arbitrary entangled states are represented in $SO(3)$. This can be done with the help of Fig. (\ref{fig5}), where $SO(3)$ is represented as a sphere of radius $\pi$ with opposite points identified. Each point of the $SO(3)$ sphere appearing in Fig  (\ref{fig5}) is an entangled state. Choosing  the origin to be state $\ket{\Psi(0)}=\alpha\ket{00}+\delta\ket{11}$, it is our initial state. Applying a rotation to one of the qubits about a given axis displaces it in the sphere  in the following way: the state performs a trajectory in the direction of the applied rotation. The angle of rotation is the distance of the final state to the the center of the sphere (initial state). In this way, a rotation about the $z$ axis performed in the first qubit, for example, creates a displacement along the $z$ axis of the sphere, leading to  state $R(z,\theta)\otimes \hat {\rm I}\ket{\Psi(0)}$. When $\theta=\pi$, the state produced can be represented in the north pole of the sphere, i.e., $i\alpha\ket{00}-i\delta\ket{11}$. Analogously, by applying such local rotations, all the other entangled states with the same degree of entanglement can be generated and represented, as depicted in Fig. (\ref{fig5}). 

For the purposes of this paper, the geometric representation described above is quite useful. We are interested in state evolutions that do not change the degree of entanglement of the initial state and are applied to  one of the qubits. Such evolutions can thus be geometrically represented in two equivalent ways. Defining, in a general pure two qubit system $\sum_{i,j}c_{ij}\ket{i_1 j_2}$, the first and second qubits by the subscripts $1,2$, let's consider that the evolution operator acts on the first qubit only. By tracing with respect to the second (fixed) qubit, the density matrix vector relative to the first (evolved) qubit describes a trajectory in the Bloch ball. This trajectory leads to a geometric and a dynamical phase. At the same time, one cannot gain information from the fiber space since, as noted above, to each point in the Bloch ball corresponds a different fiber space. Alternatively, one can choose to make a trace with respect to the evolved qubit. In this case, the density matrix vector of the reduced density matrix of the fixed qubit doesn't move in the Bloch ball. There is no geometric nor dynamic phases. However, the fiber space is always the same, and the entangled state performs an evolution in this space. This evolution leads to a total global phase  of topological nature that depends on the trajectory described by the entangled state on $SO(3)$.  In the next section, we study the details of these different phase dynamics during an entangled state  evolution.  In order to do so, we first set  notations to see more clearly the reduced density matrix vector state evolution in the Bloch ball. 


\subsection{The Reduced Density Matrix in the Bloch Ball}
Let's take as an example state 
\begin{eqnarray}\label{arb}
&&\ket{\psi_{\theta}}=\sqrt{\lambda_o}\cos{\frac{\theta}{2}}\ket{0_10_2}-\sqrt{\lambda_1}\sin{\frac{\theta}{2}}\ket{0_11_2} \\ \nonumber
&&+\sqrt{\lambda_o}\sin{\frac{\theta}{2}}\ket{1_10_2}+\sqrt{\lambda_1}\cos{\frac{\theta}{2}}\ket{1_11_2}.
\end{eqnarray}
The subscripts define the first and the second qubit, and they will be omitted in the following. The concurrence of (\ref{arb}) is $2\sqrt{\lambda_o\lambda_1}$. The angle $\theta$ has here a geometric interpretation: coordinates $X$, $Y$ and $Z$ of the state vector in the Bloch  ball are given, using the results of the previous section, by 
\begin{eqnarray}\label{coord}
&&X=(1-2\lambda_o) \sin{\theta} \nonumber \\ 
&&Y=0 \\
&&Z=(2\lambda_o-1)\cos{\theta} \nonumber.
\end{eqnarray}
The radius of the Bloch ball is $r=|2\lambda_o-1|$ and all states lying on its surface have the same degree of purity (and come from two qubit states having  the same amount of entanglement). The angle $\theta$ is the angle  the density matrix vector, represented in Fig. (\ref{fig2}),  makes with  the $z$ axis. In order to gain some geometric insight, we can start by writing the general reduced density matrix and studying some specific cases. The reduced density matrix, after tracing with respect to the second qubit, is given by 
\begin{eqnarray}
&&\hat \rho_{\theta}=(\lambda_o\cos^2\frac{\theta}{2}+\lambda_1\sin^2\frac{\theta}{2})\ket{0}\bra{0}+\nonumber \\ 
&&(\lambda_1\cos^2\frac{\theta}{2}+\lambda_o\sin^2\frac{\theta}{2}) \ket{1}\bra{1} \\ 
&&(\lambda_o-\lambda_1)\sin \theta (\ket{1}\bra{0}+\ket{0}\bra{1}. \nonumber 
\end{eqnarray}
For $\theta=0$, we have the entangled state $\ket{\Psi_{0}}=\sqrt{\lambda_o}\ket{00}+\sqrt{1-\lambda_o}\ket{11}$. The one qubit density matrix corresponding to this state is $\hat \rho_{0}=\lambda_o\ket{0}\bra{0}+(1-\lambda_o)\ket{1}\bra{1}$. The density matrix vector is thus parallel to the $z$ axis. If $\theta= \pi/2$, we have  $\ket{\Psi_{\pi/2}}=\frac{1}{\sqrt{2}}(\sqrt{\lambda_o}\ket{00}+\sqrt{1-\lambda_o}\ket{01}-\sqrt{\lambda_o}\ket{10}+\sqrt{1-\lambda_o}\ket{11} )$. The reduced density matrix  in this case is $\rho_{\pi/2}=1/2\left (\ket{0}\bra{0}+\ket{1}\bra{1}\right )+\frac{\left (\lambda_0-\lambda_1 \right )}{2}\left (\ket{0}\bra{1}+\ket{1}\bra{0}\right )$, and the vector representing it in the Bloch ball  lies in the equator.  We can see that writing (\ref{arb}) as a function of $\theta$ makes the geometric representation of the Bloch vector more direct. Let's consider now rotations performed in the first qubit of the entangled state. Such rotations also rotate the reduced density matrix vector exactly as it happens to pure states. Taking once again as the initial state  $\ket{\Psi}=\sqrt{\lambda_o}\ket{00}+\sqrt{1-\lambda_o}\ket{11}$ ($\theta=0$), its density matrix  vector lies in the $z$ direction. Rotations of the first qubit of an angle $\theta/2$ with respect to the $x$ axis via the transformation $R(\theta/2)\otimes \hat {\rm I} \ket{\Psi}$, lead to a vector making an angle $\theta$ with the $z$ axis. Rotations around other directions in space can be done  in an analogous  way. 

\section{Cyclic Phases for Pure Two Qubit States }

We can now investigate the possible origins of the phase gained by a general pure two-qubit entangled state after a cyclic evolution. We again choose, for simplicity, to consider  evolution operators acting on one qubit only, for example, the first one, as defined in (\ref{arb}). After a time interval $T$ the final state is brought to the initial one, apart from a phase factor. The initial state of the system is of the form of (\ref{initstate}). This state, as seen in the previous section, can be represented geometrically with the help of a Bloch ball of radius proportional to $\cal C$ and of $SO(3)$. The general matrix form of the evolution operator is:
\begin{equation}\label{evolop}
\hat U=e^{-i\hat H t}=    \pmatrix{
            \cos{\frac{t}{2}}-in_z \sin{\frac{t}{2}}& -in_- \sin{\frac{t}{2}}\cr
          -in_+ \sin{\frac{t}{2}} & \cos{\frac{t}{2}}+in_z \sin{\frac{t}{2}}
            },
\end{equation}
where $\hat H= \hat n \hat \sigma $, with $\hat \sigma=\hat \sigma_x \hat x+\hat \sigma_y \hat y+\hat \sigma_z \hat z$. $\hat \sigma_i$ are the Pauli matrices, $\hat n$ is the unitary vector with coordinates $\hat n= n_x\hat x+n_y\hat y+n_z\hat z$ and $\hat n_{\pm}=\hat n_x \pm i \hat n_y$. Eq. (\ref{evolop}) above represents a rotation around an axis $\hat n$ in space. The evolution operator described in (\ref{evolop}) acts only on one qubit, and the total evolution operator acting on both qubits is given by the product $\hat U \otimes \hat {\rm I}$ or $\hat {\rm I} \otimes \hat U$. The total evolution is given by a sequence of operators of the type (\ref{evolop}) acting on one qubit of the initial state (\ref{initstate}), each one of them for a fixed time interval $t$. 

In the following subsections, we study the particular case of MES and then generalize our results to states with a variable degree of entanglement. The case of MES has already been studied in \cite{EU:PRL03}, and  is of particular interest because only in this case the origin of the cyclic phase is purely topological. For non MES, we will see that geometric and dynamical effects can harmonize themselves so as to create a cyclic phase of $n\pi$, equal to the topological phase. 

\subsection{Maximally entangled states}

\begin{figure}[h]

\center

\includegraphics*[width=1.5in, keepaspectratio]{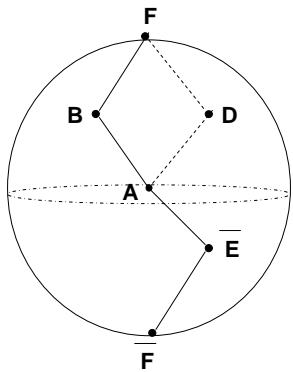}

\caption{Two possible trajectories belonging to two different homotopy groups in $SO(3)$: sequence $A \rightarrow B \rightarrow F \rightarrow D \rightarrow A$ is a plus type trajectory that does not cross the space of states orthogonal to the initial one ($A$). Sequence $A \rightarrow B \rightarrow F \rightarrow \overline E \rightarrow \overline A$ is a minus type trajectory, leading to a global phase gain of $\pi$. Note that, in this geometric view, points $A$ and $\overline A$ are both in the center of the sphere, since this point is opposite to itself. } \label{fig6}

\end{figure}

MES, as stated before, are a very particular  case in what concerns the global cyclic phase. Since the degree of entanglement  is maximal, its reduced Bloch ball is of null radius. It is  completely represented by $SO(3)$. For this reason, there is no geometric phase. We can also easily check that, using the definition for the dynamical phase $\phi_d$ (\ref{dina}),
there is no dynamical phase associated to a MES in a cyclic evolution  if one uses a sequence of operators of the form (\ref{evolop}). Nevertheless, depending on the choice of the set of operators applied to the initial state, one can either have a cyclic evolution leading to no global phase change (``plus" trajectory) or another one where a global phase of $\pi$ appears (``minus" trajectory). The existence of these two types of trajectories of MES in $SO(3)$ comes from the biconnected nature of this group \cite{LIMING:PRA04}. The ``plus" and ``minus" trajectories belong to two different homotopy classes. The difference between both classes of trajectories can be seen geometrically from the spherical representation of $SO(3)$: if one places the initial state at the origin of the $SO(3)$ sphere, the border of this sphere represents the space of states orthogonal to the initial one. They are mathematically represented by $\hat U\ket{\Psi(0)}=R(\pi, \hat n) \otimes \hat {\rm I}\ket{\Psi(0)}$. Whenever the initial state reaches this space, there is a discontinuity in the total phase. If the evolved state crosses this space an odd number of times, a total global phase of $\pi$  appears  after the cyclic evolution, defining the ``minus" type trajectories. In the ``plus" type, it crosses the same space an even number of times. The parity of the number of times the evolved states  crosses  the  space of states orthogonal to the initial one is responsible for the topological difference between trajectories  and  for the different phase gain. An example of a ``plus" and a ``minus" trajectory was studied in \cite{EU:PRL03}. They correspond to two distinct sequences of rotations applied to one of the qubits. The ``plus" trajectory is performed by a sequence of operators of the form (\ref{evolop}) generating the trajectory represented by the sequence $A \rightarrow B \rightarrow F \rightarrow D \rightarrow A$ and the ``minus" by the sequence of operators generating $A \rightarrow B \rightarrow F \rightarrow \overline E \rightarrow \overline A$, where $\overline A= -A$ and  $\overline E= -E$. $A, B, D, E, F$ are points in $SO(3)$ and both trajectories are depicted in Fig.(\ref{fig6}). Recall that rotations in $SO(3)$ are represented as displacements in the direction of $\hat n$ of a distance given by the angle of rotation $\theta$. Notice that, for the minus type of trajectory, represented pictorially by the solid line, after crossing the border of the sphere in $F$, the state reappears in the opposite side of the sphere and continues its trajectory from $\overline F$. For each part of the trajectories mentioned above, the rotation operator acts for a  time $t=2\pi/3$ and the only difference
between  the evolution operators (\ref{evolop}) performing  each part of the trajectory  is  the direction to which $\hat n$ points. These orientations $(n_x, n_y, n_z)$ read: 
\vspace{0.5cm}
\begin{center}
\begin{tabular}{|l|l|}
\hline
$A \rightarrow B$ & $ \sqrt{1/3}(-1,-1,-1)$       \\ \hline
 $B \rightarrow F$ & $\sqrt{1/3}(1,-1,-1)$   \\ \hline
 $F \rightarrow D$ & $ \sqrt{1/3}(-1,-1,1)$   \\ \hline
 $D \rightarrow A$ & $\sqrt{1/3}(-1,1,1)$   \\ \hline
$F \rightarrow \overline E$ & $ \sqrt{1/3}(-1,-1,-1)$   \\ \hline
 $\overline E \rightarrow \overline A$ & $ \sqrt{1/3}(1,-1,-1)$   \\ \hline

\end{tabular}
\vspace{0.5cm}
\end{center}
As can be seen, the axis of rotation varies in each one of the trajectories, lifting the ambiguity on the origin of the $\pi$ phase: it is manifestly a property of $SO(3)$. The case of MES subjected to a sequence of unitary evolutions of the type of Eq. (\ref{evolop}) is an example where there is no geometric nor dynamical phase. The only possible origin of the global phase gain is thus topological. We see now how these results are affected by the degree of entanglement. 

\subsection{States with an arbitrary degree of entanglement}

We study now the most general case of two-qubits with an arbitrary degree of entanglement. The limit case of a product state (that should reproduce the results obtained for two independent qubits precessing in their Bloch sphere)  appears naturally from the general results.

Consider  an arbitrary initial state of the form of (\ref{initstate}) where the first qubit only evolves, under the action of a given Hamiltonian. In principle, one should expect, when tracing with respect to the second (fixed) qubit, to have a geometric phase, since the resulting Bloch ball corresponding to the reduced density matrix  has a non null radius and the qubit describes a trajectory on it. This trajectory leads to the appearance of the expected geometric phase. At the same time, there should be a dynamical phase, and some natural questions are: what is the total phase of the non MES after one cyclic evolution? What are the contributions of the dynamical, geometrical and topological phases to this total phase? How does it depend on entanglement? A  way to start to answer this question is to apply to the non MES the two sequences of transformations that were performed to the MES  in the previous section and calculate the total phase with the help of (\ref{pancha}). Taking as initial state, for example,
\begin{equation}\label{aiai}
\ket{\Psi(0)}=\sqrt{\lambda_o}\ket{00}+\sqrt{1-\lambda_o}\ket{11}
\end{equation}
there are two equivalent scenarios, that must lead to the same final result since they depend on the pure entangled state only: tracing (\ref{aiai}) with respect to the evolved or the fixed qubit. If one traces out the evolved qubit, all the phases should come  from the trajectory of the non MES in $SO(3)$. By tracing with respect to the second (fixed) qubit, the phase information must come from the reduced density matrix vector's trajectory in the Bloch ball and the dynamical phase. Let's consider first the case where one traces out the evolved qubit. Defining (\ref{aiai}) as the center of the sphere representing $SO(3)$, we have that, after applying the same sequence of rotations discussed in the previous subsection, there are still two types of trajectories (plus and minus) in $SO(3)$ that are represented exactly in the same way as Fig.(\ref{fig6}). In general, we therefore have that {\it the total phase after the cyclic evolution is  the same as for the MES case regardless of the value of $\lambda_o$}.
\begin{figure}[h]

\center

\includegraphics*[width=3.2in, keepaspectratio]{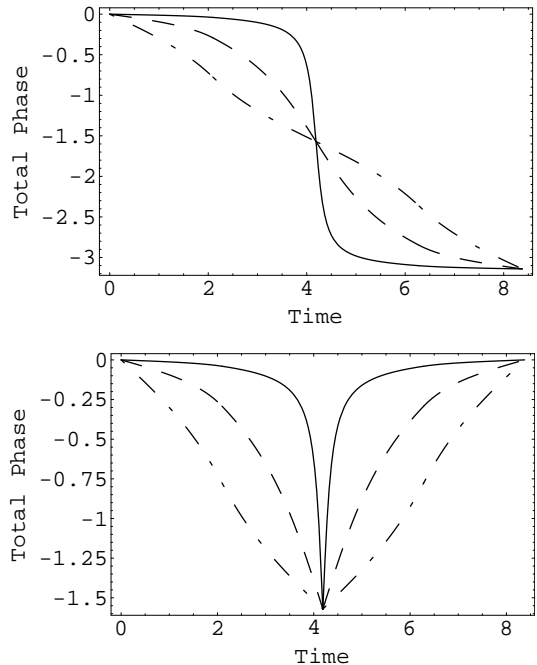}

\caption{Total phase for trajectories  ``minus" (above) and ``plus" (below) for three values of the parameter $\lambda_o$: 0.3 (dot-dashed line), 0.4 (dashed line) and 0.48 (continuous line). The curves become more sharp and tend to discontinuity as the value of $\lambda_o$ increases. For $\lambda_o=0,1$ (product states), it is given by a straight line, as expected.} \label{fig7}

\end{figure}

\begin{figure}[h]

\center

\includegraphics[width=2in, keepaspectratio]{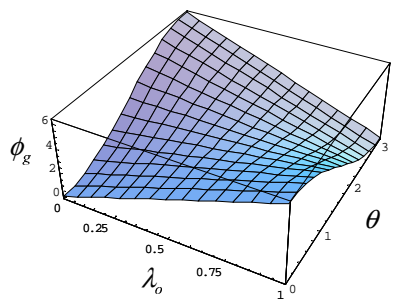}

\caption{Geometric phase for a cyclic evolution around a fixed axis as a function of the angle $\theta$ and of parameter $\lambda_o$, proportional to the degree of entanglement. } \label{fig8}

\end{figure}

Indeed, the existence of the two types of trajectory doesn't depend on the degree of entanglement. The total phase can be calculated from (\ref{pancha}). Fig. (\ref{fig7}) shows it for three possible values of $\lambda_o=0.3, 0.4$ and $0.48$ for both types of trajectory.  We see that for the three values of $\lambda_o$ one can, effectively, define two types of trajectory. However, the closer we are to MES ($\lambda_o=0.5$), the more abrupt the phase gain becomes, until it reaches a discontinuity for $\lambda_o=0.5$. In this limit case, the phase is either zero or $\pi$ during the whole trajectory and the discontinuity represents the gain of the topological phase. As seen before, it is a consequence of  state's $\ket{\Psi(t)}$ crossing the border of the $SO(3)$ sphere in the middle of the trajectory ($t=4\pi/3$). Fig(\ref{fig7}) shows that the same happens for non MES.

Let's consider now the same trajectories but trace with respect to the fixed qubit. This results on the reduced density matrix of the evolved qubit. This density matrix performs a trajectory in the Bloch ball. As mentioned above, the existence of the two types of trajectory is a mathematical result, valid for the entangled pure state, that must be reproduced in the two situations: tracing with respect to the evolved or the fixed qubit. By tracing with respect to the fixed qubit, one can calculate the global phase from its definition (\ref{pancha}). In order to do so, we should first calculate the scalar product:
\begin{equation}\label{SP}
SP=\cos\frac{t}{2}+i(Xn_x(t)+Yn_y(t)+Zn_z(t))\sin\frac{t}{2}.
\end{equation}
In the expression above,  $X, Y$ and $Z$ are the coordinates of the Bloch vector of the reduced density matrix and the vector $\hat n$ was taken as a function of time, as proposed in \cite{LIMING:PRA04}. In the example studied, ${\hat n}(t)$  suffers discontinuous changes corresponding to the different axes of rotation. We see that Eq.(\ref{SP}) is completely determined by the coordinates of the reduced density matrix vector in the Bloch ball and the same result could be obtained by calculating the mixed state total phase $\phi_t^M={\rm  arg Tr}[\hat U\hat \rho]$. One can  see that for the particular initial state (\ref{aiai}) and the sequence of rotations under consideration, the scalar product  never reaches the value of zero, so there is no discontinuity in the phase. This result has already been seen in Fig.(\ref{fig7})  as being a consequence of the trajectory of the entangled state in $SO(3)$. However, Eq.(\ref{SP}) shows clearly that the same result can be obtained from information on the Hamiltonian and the reduced density matrix only. Eq.(\ref{SP}) also shows that a total $n\pi$ phase still appears. This result can be seen  as being a consequence of the parity of the number of times the initial state crosses the border of the $SO(3)$ sphere or of the area of the trajectory of the reduced density matrix in the Bloch ball combined to the dynamical phase. 

\begin{figure}[h]

\center

\includegraphics[width=2in, keepaspectratio]{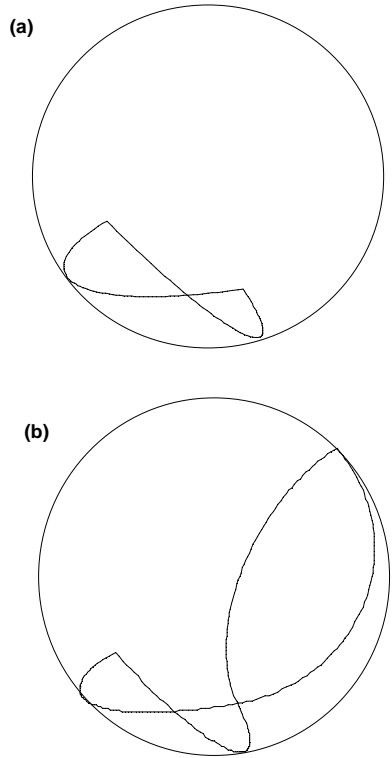}

\caption{Trajectories in the Bloch ball for the two possible sets of rotations  (a) $A \rightarrow B \rightarrow F \rightarrow \overline E \rightarrow \overline A$ (minus trajectory), and (b) $A \rightarrow B \rightarrow F \rightarrow D \rightarrow A$ (plus trajectory).  } \label{fig9}

\end{figure}

We can now calculate the contribution coming from the dynamical and the geometric phases of the reduced density matrix. The calculation of the dynamical phase is performed from its definition (\ref{dina}). It is given by the integral
\begin{equation}\label{dinamic}
\frac{1}{2}\int^t_0(n_x(t)X+n_y(t)Y+n_z(t)Z)dt, 
\end{equation}
that can be seen as the integral of the scalar product between the density matrix vector and the direction of rotation $\hat n(t)$. For this specific example, the dynamical phase is still zero. It is thus the geometric phase that explains the existence of the two types of trajectories. The two different sequences of evolution operators lead to different trajectories in the Bloch ball that are independent of the degree of entanglement. They are shown in Fig.(\ref{fig9}). The area of each one of this trajectories can be calculated, and by weighting them using the techniques of Section II, one finds the two possible global phases, $0$ or $\pi$. 

We see that either one can obtain the total global phase of a non MES entangled state from the topological properties of $SO(3)$, or by geometric properties of a reduced density matrix. This means that by looking at a subsystem of the entangled state gives us information about its global phase.

As a general rule, the scalar product (\ref{SP}) and the dynamical phase (\ref{dinamic}) can  tell us all about the types of phases appearing in an evolution. To illustrate that, let's study a simpler Hamiltonian, describing a rotation around only one fixed axis, say, the $z$ axis. For a general state of the type (\ref{arb}), (\ref{SP}) depends on $\theta$ as $SP=\cos\frac{t}{2}+i(2\lambda_o-1)\cos\theta\sin\frac{t}{2}$. The dynamic phase for this case is $\phi_d=\pi(2\lambda_o-1)\cos{\theta}$. Notice that, apart from a scaling factor depending on the degree of entanglement $(2\lambda_o-1)$, it has the same form as for pure states. The cyclic geometric phase  also depends on $\theta$ and $\lambda_o$. It is given by  $\phi_g=\pi+\pi(1-2\lambda_o) \cos{\theta}$,  plotted in Fig. (\ref{fig8}). The connection of this phase to the pure state case is not direct. However, some limit cases are reproduced. First, notice that for $\lambda_o=1/2$ (MES), $\phi_g$ and $\phi_d$ vanish, as expected. Also, for separable states ($\lambda_o=0,1$), $\phi_g$ oscillates with $\theta$ as in the pure state example discussed in Section II. At the same time, in this limit, the total phase is a linear function of time, as one should expect. The dependence of the phases on $\theta$ of $\phi_g$ is also interesting: in the limit $\theta=\pi/2$, $\phi_d$ and $\phi_g$ show that the total global phase is of geometric nature only, a result analogous to the one found for one qubit states. However, for $\theta=0$, the analogy with the one qubit case breaks down. The geometric phase is given by $\phi_g=2\pi\lambda_1$ in contrast to zero for the case of one qubit. Nevertheless, there is no contradiction, since in the discussed  case of mixed states the dynamical phase doesn't account for the total phase. It assumes the value of $\pi(2\lambda_o-1)$, so both add together to $\pi$, as expected. It is also easy to verify that for an arbitrary value of $\theta$, $\phi_g+\phi_d=\pi$, while the proportions of geometrical and dynamical  phases are determined by the angle $\theta$. By looking now again at the case where the first qubit is traced away,  for an evolution representing a rotation around one fixed axis in space, all trajectories cross the border of the $SO(3)$ sphere, accounting for a global phase of $\pi$. This is why we must observe $\phi_g+\phi_d=\pi$.

The discussion above generalizes the results  of \cite{EU:PRL03} and gives a geometric  interpretation of the nature of the cyclic phases for entangled two-qubit states with an arbitrary degree of entanglement. This is done with the help of geometric phases represented in the Bloch ball for density matrices. As for the case of MES, the results presented can be, in principle, tested experimentally. To do so, one needs to throw the non MES in an interferometer, so that the global phase can become a relative phase. A possible way to do that is using photon interference, as proposed in \cite{EU:PRL03}. Recently, other proposals have appeared aiming to show the same effects in a system of entangled  spin pairs \cite{LIMING:PRA04}. Both proposals need slight modifications in order to account for the phase effects described in this paper. This modification concerns only the initial state, that now can have an arbitrary degree of entanglement. Taking the example of polarized twin photons, it has already been demonstrated that polarisation entangled photon pairs with an arbitrary degree of entanglement can be generated \cite{KWIAT}. As shown above, the same set of transformations performed in MES lead to the two different types of trajectories also for non MES. This means that the same set of wave plates used in \cite{EU:PRL03} can be employed to test the predicted results.

\section{Conclusions}

To conclude, the different types of phases appearing in a cyclic evolution of a pure bipartite entangled state were studied. This is done from a geometric point of view. It is shown that the interplay between topological, dynamical and geometrical phases always lead to a global phase of $n\pi$. The parity of $n$, can have different origins, depending on how one choses to geometrically represent the pure entangled state: by considering evolution operators acting on one qubit only and tracing with respect to the fixed qubit,  trajectories of the type plus and minus are a consequence of geometric and dynamical phases of the evolved qubit. Alternatively, still supposing that only one qubit only evolves and tracing it out, the parity of the total phase is a consequence of the number of times an initial state, placed at the origin of $SO(3)$, crosses the border of this sphere. As a consequence, it is shown that mixed states coming from the trace of a pure entangled state with respect to one of the particles may play a crucial role in describing the total phase gain of the pure entangled states itself. The author wishes to thank  R. Mosseri for inspiring discussions and I. Lorger\'e for a critical reading of the manuscript.






\begin{thebibliography}{99}


\bibitem{PAN} S. Pancharatnam, Proc. Indian Acad. Sci. Sec. A {\bf 44}, 247 (1956).
\bibitem{BERRY} M. V. Berry, Proc. R. Soc. London A {\bf 392}, 45 (1984).
\bibitem{BOOK} C. Nash and S. Sen, {\it Topology and Geometry for Physicists}, Academic Press (1983).
\bibitem{SJOQVIST:PRA00} E. Sj\"oqvist, Phys. Rev. A {\bf 62}, 022109 (2000).

\bibitem{KWIAT:PRL91} P. G. Kwiat and R. Chiao, Phys. Rev. Lett. {\bf 66}, 588 (1991). 
\bibitem{COND}I. Fuentes-Guridi, J. Pachos, S. Bose, V. Vedral, and S. Choi, Phys. Rev. A {\bf 66}, 022102 (2002).
\bibitem{AHARONOV:PRL00} Y. Aharonov and B. Reznik, Phys. Rev. Lett. {\bf 84}, 4790 (2000).
\bibitem{MARCELO} A. Carollo {\it et al.} Phys. Rev. Lett. {\bf 90}, 160402 (2003); A. Carollo {\it et al.} Phys. Rev. Lett. {\bf 92}, 020402 (2004).  
\bibitem{DUAN:SCIENCE01} L. M. Duan, J. I. Cirac and P. Zoller, Science {\bf 292}, 1695 (2001); G. Milburn, S. Schneider and D. F. James, Fortschr. Physik {\bf 48} 801 (2000); X. Wang, A. Sorensen, K. Molmer, Phys. Rev. Lett. {bf 86}, 3907 (2001). 

\bibitem{WINE:NATURE03} D. Leibfried {\it et al.} Nature {\bf 422}, 412-415 (2003); J. A. Jones {\it et al.}, Nature {\bf 403}, 869 (2000).
\bibitem{KITAEV:ANNALS03} A. Kitaev, Annals Phys. 2-30, (2003).

\bibitem{EU:PRL03} P. Milman and R. Mosseri, Phys. Rev. Lett. {\bf 90 }, 230403 (2003).

\bibitem{MOSSERI:JPA01} R. Mosseri and R. Dandoloff, J. Phys. A: Math. Gen {\bf 34} (2001).
\bibitem{NEUTRONS} H. Rauch {\it et al.}, Phys. Lett {\bf 54 A}, 425 (1975); S. A. Werner {\it et al.}, Phys. Rev. Lett. {bf 35}, 1053 (1975). 
\bibitem{URBANKE:AJP91} H. Urbanke, Am. J. Phys.  {\bf 59}, 53 (1991).

\bibitem{OI:PRL00} E. Sj\"oqvist {\it et al.} Phys. Rev. Lett. {\bf 85}, 2845 (2000).
\bibitem{OI:PRL03} M. Ericsson {\it et al.} Phys. Rev. Lett. {\bf 91}, 090405 (2003)

\bibitem{TONG:PRA03} D. M. Tong {\it et al. } Phys. Rev. A {\bf 68}, 022106 (2003).

\bibitem{LIMING:PRA04} W. LiMing, Z. L. Tang and C. J. Tao, Phys. Rev. A {\bf 69}, 064301 (2004).
\bibitem{WOOTERS:PRL98} W. K. Wootters, Phys. Rev. Lett. {\bf 80}, 2245 (1998). 
    
 

\bibitem{KWIAT} K. Mattle {\it et al.}, Phys. Rev. Lett. {\bf 76}, 4656 (1996); D. Bouwemeester {\it al.}, Phys. Rev. Lett. {\bf 82}, 1345 (1999); M. Fran\c ca Santos {\it et al.}  Phys. Rev. A {\bf 023804} (2002).     

  
\end{thebibliography}
\end{document}